# Guaranteed Minimum Rank Approximation from Linear Observations by Nuclear Norm Minimization with an Ellipsoidal Constraint

Kiryung Lee and Yoram Bresler

October 29, 2018

**Abstract**

The rank minimization problem is to find the lowest-rank matrix in a given set. Nuclear norm minimization has been proposed as an convex relaxation of rank minimization. Recht, Fazel, and Parrilo have shown that nuclear norm minimization subject to an affine constraint is equivalent to rank minimization under a certain condition given in terms of the rank-restricted isometry property. However, in the presence of measurement noise, or with only approximately low rank generative model, the appropriate constraint set is an ellipsoid rather than an affine space. There exist polynomial-time algorithms to solve the nuclear norm minimization with an ellipsoidal constraint, but no performance guarantee has been shown for these algorithms. In this paper, we derive such an explicit performance guarantee, bounding the error in the approximate solution provided by nuclear norm minimization with an ellipsoidal constraint.

# 1 Introduction

The rank minimization problem is to find the lowest rank matrix in a given set $\mathcal{C}$ [FHB01], i.e.,

$$\begin{array}{ll} \min_{X \in \mathbb{C}^{m\times n}} & \mathrm{rank}(X) \\ \text{subject to} & X \in \mathcal{C}. \end{array} \tag{1}$$

In particular, there are applications such as matrix completion and minimum order system identification [1] that require the reconstruction of low-rank matrix $X \in \mathbb{C}^{m\times n}$ from the linear measurement $b = \mathcal{A}X \in \mathbb{C}^p$ obtained with a given linear operator $\mathcal{A} : \mathbb{C}^{m\times n} \to \mathbb{C}^p$. In this case, the set $\mathcal{C}$ is given as an affine space by $\mathcal{C} = \{X : \mathcal{A}X = b\}$ and we are solving an inverse problem $\mathcal{A}X = b$ for $X$ with the *a priori* information that the true solution is a low-rank matrix.

In general, rank minimization is a difficult non-convex optimization problem and no polynomial time algorithm has been proposed to date. Nuclear norm minimization [FHB01] is a convex relaxation of the rank minimization problem with a convex set $\mathcal{C}$. Recht, Fazel, and Parrilo derived a performance guarantee for nuclear norm minimization with an affine constraint [RFP07]. A sufficient condition for the performance guarantee is given in terms of the rank-restricted isometry property of the linear operator $\mathcal{A}$. Roughly, when $\mathcal{A}$ is nearly an isometry for low-rank matrices, rank minimization is equivalent to nuclear norm minimization and hence can be solved in polynomial time.

However, in some applications of rank minimization such as minimum order system approximation, reduced order controller design, and the Euclidean distance matrix problem [LLR95], the inverse problem $\mathcal{A}X = b$ with given linear operator $\mathcal{A}$ and measurement $b$ may not admit a low-rank solution. For example, in minimum order system approximation, the given system cannot be described by a low-rank matrix but can be well approximated by one. In this case, the minimum rank of solutions to $\mathcal{A}X = b$, which is given by $\min_X\{\text{rank}(X) : \mathcal{A}X = b\}$, can be higher than the desired target rank. Another possibility is that there is additive noise in the measurements. Again the inverse problem $\mathcal{A}X = b$ may not admit a low-rank solution. Instead, in order to find a low-rank approximate solution whose rank is lower than the target value required by the application, the set $\mathcal{C}$ can be modified to an ellipsoid given by

$$\mathcal{C} = \{X : \|\mathcal{A}X - b\|_2 \leqslant \epsilon\}. \tag{2}$$

[1] For more applications of rank minimization with an affine constraint, see [RFP07], [Faz02], and the reference therein.

The resulting rank minimization problem defined by (1) and (2) is hard, and its nuclear norm convex relaxation can be used to obtain approximate solutions. In fact, there exist polynomial-time algorithms to solve the nuclear norm minimization problem with an ellipsoidal constraint (e.g. [FHB01], [CCS08]). However, while empirically effective, a theoretical performance guarantee for those algorithms has been missing. Our goal in this paper is to close this gap in theory. We are motivated by the analogy established by Recht, Fazel, and Parrilo [RFP07] between the rank minimization problem and $\ell_0$ norm minimization, or equivalently compressed sensing, for the affine constraint case. This analogy extends to the convex relaxations of these problems, nuclear norm minimization and $\ell_1$ norm minimization, respectively [RFP07].

For the affine constraint case, Candes and Tao [CT05] have given a sufficient condition for the equivalence of $\ell_0$ norm minimization to its $\ell_1$ relaxation (or basis pursuit) in the sense that both problems admit the same and unique solution. The condition is given in terms of the sparsity-restricted isometry property of the sensing matrix. For the ellipsoidal constraint case, also known as the noisy and compressible signal case, Candes extended the performance guarantee of $\ell_1$ norm minimization showing that the error in the sparse approximate solution is bounded by a weighted sum of the best sparse approximation error of the true solution and a bound on the energy of the noise in the measurement [Can08]. An analogous performance guarantee for nuclear norm minimization with an ellipsoidal constraint has not been available to date.

In this paper, we seek the relation between the rank minimization problem with an ellipsoidal constraint and its convex relaxation. Basically, we use an analogue of the approach by Candes for $\ell_1$ norm minimization [Can08]. The extended performance guarantee is given in terms of the rank-restricted isometry property and bounds the error in the low-rank approximate solution by a weighted sum of the error in the best low-rank approximation of the true solution, and a bound on the energy of the measurement noise.

# 2 Performance Guarantee

Consider two Hilbert spaces $\mathbb{C}^{m\times n}$ and $\mathbb{C}^p$. For $X, Y \in \mathbb{C}^{m\times n}$, the inner product is defined by $\langle X, Y\rangle_{\mathbb{C}^{m\times n}} = \mathrm{Tr}(Y^H X)$, where $Y^H$ denotes the Hermitian transpose of $Y$. Then the induced Hilbert-Schmidt norm on $\mathbb{C}^{m\times n}$ is the Frobenius norm and will be denoted by $\|\cdot\|_F$. For $x, y \in \mathbb{C}^p$, the inner product is defined by $\langle x, y\rangle_{\mathbb{C}^p} = y^H x$. Then the induced Hilbert-Schmidt norm of $\mathbb{C}^p$ is the Euclidean norm and will be denoted by $\|\cdot\|_2$.

The setting for the low-rank matrix recovery and approximation problem is the following. The measurement (with perturbation) of an unknown matrix $X \in \mathbb{C}^{m\times n}$ is given as $b = \mathcal{A}X + \nu$ with $\|\nu\|_2 \leqslant \epsilon$ where $\mathcal{A} : \mathbb{C}^{m\times n} \to \mathbb{C}^p$ is a given linear operator. The inverse problem is to recover matrix $X$, which is considered as an unknown true solution, with the side information that $X$ has low rank or can can be well-approximated by such a matrix. Accordingly, the problem may be formulated as in (1), with the ellipsoidal constraint (2). The convex relaxation of this problem is the nuclear norm minimization problem

$$\text{P:} \quad \begin{array}{ll} \min\limits_{X \in \mathbb{C}^{m\times n}} & \|X\|_* \\ \text{subject to} & \|\mathcal{A}X - b\|_2 \leqslant \epsilon, \end{array}$$

where $\|X\|_*$ denotes the nuclear norm, which is the sum of the singular values of $X$. Problem P admits a low-rank solution that is a low-rank approximate solution to the original inverse problem. The quality of this approximate solution can be guaranteed subject to a condition on the *rank-restricted isometry constant.*

Given a linear operator $\mathcal{A} : \mathbb{C}^{m\times n} \to \mathbb{C}^p$, the rank-restricted isometry constant $\delta_r(\mathcal{A})$ is defined as the minimum constant that satisfies

$$(1 - \delta_r(\mathcal{A})) \|X\|_F^2 \leqslant \|\mathcal{A}X\|_2^2 \leqslant (1 + \delta_r(\mathcal{A})) \|X\|_F^2 , \tag{3}$$

for all $X \in \mathbb{C}^{m\times n}$ with $\text{rank}(X) \leqslant r$. [2]

**Theorem 2.1** *Let $X^\star$ be the solution to P. If $\mathcal{A}$ has the rank-restricted isometry constant $\delta_{3r}(\mathcal{A}) < 1/(1 + 4/\sqrt{3})$, then*

$$\|X^\star - X\|_F \leqslant K_0 \|X - X_r\|_F + K_1\epsilon, \tag{4}$$

*where $X_r$ denotes the best rank-$r$ approximation of $X$ given by*

$$X_r \triangleq \arg \min_{Z\in\mathbb{C}^{m\times n}} \{\|X - Z\|_F : \text{rank}(Z) \leqslant r\}.$$

[2] The definition of the rank-restricted isometry property is slightly different from that in [RFP07] in the sense that the norms in the inequality are squared in our definition. This is done for the consistency with the sparsity-restricted isometry for $\ell_0$ norm minimization [Can08].

*The constants $K_0$ and $K_1$ are given as*

$$\begin{aligned} K_0 &= \left(\frac{4\sqrt{2}}{\sqrt{3}}\right)\frac{(1+(\sqrt{2}-1)\delta_{3r}(\mathcal{A}))}{1-(1+4/\sqrt{3})\delta_{3r}(\mathcal{A})} \\ K_1 &= \left(\frac{\sqrt{3}+2\sqrt{2}}{\sqrt{3}}\right)\frac{2\sqrt{1+\delta_{3r}(\mathcal{A})}}{1-(1+4/\sqrt{3})\delta_{3r}(\mathcal{A})}, \end{aligned}$$

*respectively.*

The two terms $\|X - X_r\|_F$ and $\epsilon$ in the bound of (4) reflect the compressibility of matrix $X$, and the strength of the measurement noise, respectively. In general $X$ may not be exactly low-rank with $\text{rank}(X) \leqslant r$ but $X$ admits a good low-rank approximation with small $\|X - X_r\|_F$. The measurements are also subject to a perturbation. These imperfections cause an error in the low-rank approximate solution obtained by the nuclear norm minimization. However, the gain of each term is explicitly bounded by a constant determined by the rank-restricted isometry constant. In particular, when $\epsilon = 0$ and $\|X - X_r\|_F = 0$, the solution obtained by the nuclear norm minimization coincides with the true solution $X$. Furthermore, an immediate corollary of Theorem 2.1 states that rank minimization with an ellipsoidal constraint can be solved by nuclear norm minimization in polynomial time in the sense that the distance between the solutions obtained by rank minimization and nuclear norm minimization is bounded as a linear function of $\|X - X_r\|_F$ and $\epsilon$.

# 3 Proof of Performance Guarantee

We first note the *rank-restricted orthogonality property* that follows from the rank-restricted isometry property. The following Proposition is an extension of Lemma 2.1 in [Can08] for the vector case to the matrix case.

**Definition 3.1** *Given a set $\Psi = \{\psi_1, \ldots, \psi_{|\Psi|}\} \subset \mathbb{C}^{m \times n}$, define a linear oper-*

*ator* $L_\Psi : \mathbb{C}^{|\Psi|} \to \mathbb{C}^{m\times n}$ *by*

$$L_\Psi \alpha = \sum_{k=1}^{|\Psi|} \alpha_k \psi_k, \quad \forall \alpha \in \mathbb{C}^{|\Psi|}. \tag{5}$$

It follows from (5) that the adjoint operator $L_\Psi^* : \mathbb{C}^{m\times n} \to \mathbb{C}^{|\Psi|}$ is given by

$$(L_\Psi^* X)_k = \langle X, \psi_k \rangle_{\mathbb{C}^{m\times n}}, \quad \forall k = 1, \ldots, |\Psi|, \ \forall X \in \mathbb{C}^{m\times n}. \tag{6}$$

Note that for $\mathcal{A} : \mathbb{C}^{m\times n} \to \mathbb{C}^p$ the operator composition $\mathcal{A}L_\Psi : \mathbb{C}^{|\Psi|} \to \mathbb{C}^p$ admits a matrix representation. Its pseudo-inverse is denoted by $[\mathcal{A}L_\Psi]^\dagger$.

**Remark 3.2** *If the elements in* $\Psi$ *are pairwise orthogonal and normalized, then* $L_\Psi$ *is an isometry and the projection* $P_\Psi$ *onto* $\mathrm{span}(\Psi)$ *is given by* $P_\Psi = L_\Psi L_\Psi^*$. *If* $\Psi$ *is a set of rank-one matrices, then* $\mathrm{rank}(L_\Psi \alpha) \leqslant |\Psi|$ *for all* $\alpha \in \mathbb{C}^{|\Psi|}$.

**Proposition 3.3** *Suppose that linear operator* $\mathcal{A} : \mathbb{C}^{m\times n} \to \mathbb{C}^p$ *has the rank-restricted isometry constant* $\delta_r(\mathcal{A})$. *For* $X \in \mathbb{C}^{m\times n}$, *let* $X = \sum_{j=1}^{\mathrm{rank}(X)} \sigma_j \psi_j$ *denote the singular value decomposition of* $X$ *where* $\psi_j$ *is a rank-one unit-norm matrix obtained by the outer product of left and right singular vectors corresponding to the* $j$*-th singular value* $\sigma_j$ *for* $j = 1, \ldots, \mathrm{rank}(X)$. *Similarly, for* $Y \in \mathbb{C}^{m\times n}$, *let* $Y = \sum_{j=1}^{\mathrm{rank}(Y)} \sigma_j' \psi_j'$ *denote the singular value decomposition of* $Y$. *If* $\langle \psi_j, \psi_k' \rangle_{\mathbb{C}^{m\times n}} = 0$ *for all* $j = 1, \ldots, \mathrm{rank}(X)$ *and* $k = 1, \ldots, \mathrm{rank}(Y)$ *and* $\mathrm{rank}(X) + \mathrm{rank}(Y) \leqslant r$, *then*

$$|\langle \mathcal{A}X, \mathcal{A}Y \rangle_{\mathbb{C}^p}| \leqslant \delta_r(\mathcal{A}) \, \|X\|_F \, \|Y\|_F \, . \tag{7}$$

**Proof** Let $\Psi = \{\psi_j\}_{j=1}^{\mathrm{rank}(X)}$, $\Psi' = \{\psi_j'\}_{j=1}^{\mathrm{rank}(Y)}$, and $\widetilde{\Psi} = \Psi \cup \Psi'$. Then $L_\Psi$, $L_{\Psi'}$,

and $L_{\widetilde{\Psi}}$ are isometries. Therefore, together with the R-RIP of $\mathcal{A}$, it follows that

$$1-\delta_r(\mathcal{A}) \leqslant \sigma_{\min}(L_{\widetilde{\Psi}}^* \mathcal{A}^* \mathcal{A} L_{\widetilde{\Psi}}) \leqslant \sigma_{\max}(L_{\widetilde{\Psi}}^* \mathcal{A}^* \mathcal{A} L_{\widetilde{\Psi}}) \leqslant 1+\delta_r(\mathcal{A}).$$

Note that $L_{\Psi'}^* \mathcal{A}^* \mathcal{A} L_{\Psi}$ is an off-diagonal submatrix of $L_{\widetilde{\Psi}}^* \mathcal{A}^* \mathcal{A} L_{\widetilde{\Psi}}$, and therefore also of $L_{\widetilde{\Psi}}^* \mathcal{A}^* \mathcal{A} L_{\widetilde{\Psi}} - I_d$, where $I_d$ is the identity matrix of compatible size. Hence

$$\begin{aligned}
\sigma_{\max}(L_{\Psi'}^* \mathcal{A}^* \mathcal{A} L_{\Psi}) &\leqslant \sigma_{\max}(L_{\widetilde{\Psi}}^* \mathcal{A}^* \mathcal{A} L_{\widetilde{\Psi}} - I_d) \\
&\leqslant \max\{(1+\delta_r(\mathcal{A}))-1, 1-(1-\delta_r(\mathcal{A}))\} \\
&= \delta_r(\mathcal{A}).
\end{aligned}$$

Noting that $X = P_\Psi X = L_\Psi L_\Psi^* X$ and $Y = P_{\Psi'} Y = L_{\Psi'} L_{\Psi'}^* Y$, it follows

$$\begin{aligned}
|\langle \mathcal{A}X, \mathcal{A}Y \rangle_{\mathbb{C}^p}| &= |\langle \mathcal{A} L_\Psi L_\Psi^* X, \mathcal{A} L_{\Psi'} L_{\Psi'}^* Y \rangle_{\mathbb{C}^p}| \\
&= |\langle [L_{\Psi'}^* \mathcal{A}^* \mathcal{A} L_\Psi] L_\Psi^* X, L_{\Psi'}^* Y \rangle_{\mathbb{C}^{|\Psi'|}}| \\
&\leqslant \sigma_{\max}(L_{\Psi'} \mathcal{A}^* \mathcal{A} L_\Psi) \left\| L_\Psi^* X \right\|_2 \left\| L_{\Psi'} Y \right\|_2 \\
&\leqslant \delta_r(\mathcal{A}) \left\| L_\Psi^* X \right\|_2 \left\| L_{\Psi'}^* Y \right\|_2 \\
&= \delta_r(\mathcal{A}) \left\| L_\Psi L_\Psi^* X \right\|_F \left\| L_{\Psi'} L_{\Psi'}^* Y \right\|_F \\
&= \delta_r(\mathcal{A}) \left\| X \right\|_F \left\| Y \right\|_F .
\end{aligned}$$

∎

Next we note certain properties of the nuclear norm.

**Lemma 3.4** *Let $X, Y \in \mathbb{C}^{m\times n}$. Then $\|X+Y\|_* = \|X\|_* + \|Y\|_*$ if and only if $X$ and $Y$ are simultaneously diagonalizable into nonnegative matrices.*

**Proof** Let $\Gamma_m$ and $\Gamma_n$ denote the sets of unitary matrices in $\mathbb{C}^{m\times m}$ and $\mathbb{C}^{n\times n}$,

respectively. By the variational principle,

$$\|X\|_* = \max_{U\in\Gamma_m, V\in\Gamma_n} \mathrm{Tr}(U^H X V). \tag{8}$$

Moreover, $(U,V)$ is a maximizer of (8) if and only if $U^H XV$ is a diagonal matrix where the diagonal entries are singular values of $X$. Equation (8) implies

$$\begin{aligned}
\|X+Y\|_* &= \max_{U\in\Gamma_m, V\in\Gamma_n} \mathrm{Tr}(U^H (X+Y) V) && (9)\\
&\leqslant \max_{U\in\Gamma_m, V\in\Gamma_n} \mathrm{Tr}(U^H X V) + \max_{U\in\Gamma_m, V\in\Gamma_n} \mathrm{Tr}(U^H Y V) && (10)\\
&= \|X\|_* + \|Y\|_* .
\end{aligned}$$

Let $(U_0, V_0)$ denotes a maximizer of (9). The equality in (10) holds if and only if both $U_0^H XV_0$ and $U_0^H YV_0$ are diagonal matrices and the diagonal entries of $U_0^H XV_0$ and $U_0^H YV_0$ correspond to the singular values of $X$ and $Y$, respectively. Noting that the singular values are nonnegative completes the proof. ∎

**Corollary 3.5** (Lemma 2.3 in [RFP07]) *Let $X, Y \in \mathbb{C}^{m\times n}$. If $XY^H = 0$ and $X^H Y = 0$, then*

$$\|X+Y\|_* = \|X\|_* + \|Y\|_* .$$

**Proof** Let $X = U_1\Sigma_1 V_1^H$ and $Y = U_2\Sigma_2 V_2^H$ denote the singular value decompositions of $X$ and $Y$, respectively. The assumption implies that $V_1^H V_2 = 0$ and $U_1^H U_2 = 0$. Let $U = [\ U_1 \quad U_2\ ]$ and $V = [\ V_1 \quad V_2\ ]$. By concatenating orthonomal columns to $U$ and $V$, we construct unitary matrices $\widetilde{U}$ and $\widetilde{V}$ which have $U$ and $V$ as their submatrices, respectively. Then $(\widetilde{U}, \widetilde{V})$ simultaneously diagonalize $X$ and $Y$ into nonnegative matrices and hence the result follows by Lemma 3.4. ∎

**Lemma 3.6** *Let* $X \in \mathbb{C}^{m\times n}$*, and suppose* $\operatorname{rank}(X) \leqslant r$*. Then*

$$\|X\|_F \leqslant \|X\|_* \leqslant r^{1/2}\,\|X\|_F\,.$$

**Proof** Let $(\sigma_k)_{k=1}^r$ denotes the singular values of $X$ in decreasing order where $r$ is the rank of $X$. Since $\sigma_k \geqslant 0$ for all $k = 1, \ldots, r$,

$$\sqrt{\sum_{k=1}^r \sigma_k^2} \leqslant \sum_{k=1}^r \sigma_k \leqslant r^{1/2}\sqrt{\sum_{k=1}^r \sigma_k^2},$$

where the second inequality follows from the Cauchy-Schwartz inequality. Noting $\|X\|_F^2 = \sum_{k=1}^r \sigma_k^2$ and $\|X\|_* = \sum_{k=1}^r \sigma_k$ completes the proof. ∎

**Proof of Theorem 2.1**

Let $X = U\Sigma V^H$ denote the full singular value decomposition of $X$, where $U \in \mathbb{C}^{m\times m}, \Sigma \in \mathbb{C}^{m\times n}, V \in \mathbb{C}^{n\times n}$. Let $u_k$, $v_k$ denote the $k$-th column of $U$ and $V$, respectively. Then define four projection operators in terms of the $u_k$'s and $v_k$'s:

$$\begin{aligned}
P_1 Z &= \sum_{j=1}^r \sum_{k=1}^r \langle Z, u_j v_k^H\rangle_{\mathbb{C}^{m\times n}} u_j v_k^H \\
P_2 Z &= \sum_{j=r+1}^m \sum_{k=1}^r \langle Z, u_j v_k^H\rangle_{\mathbb{C}^{m\times n}} u_j v_k^H \\
P_3 Z &= \sum_{j=1}^r \sum_{k=r+1}^n \langle Z, u_j v_k^H\rangle_{\mathbb{C}^{m\times n}} u_j v_k^H \\
P_4 Z &= \sum_{j=r+1}^m \sum_{k=r+1}^n \langle Z, u_j v_k^H\rangle_{\mathbb{C}^{m\times n}} u_j v_k^H\,.
\end{aligned}$$

Obviously, $P_1 + P_2 + P_3 + P_4 = I$, where $I$ is the identity operator on $\mathbb{C}^{m\times n}$. Also, $X_r = P_1 X$. By construction, $(P_1 Z)(P_4 Z)^H = 0$ and $(P_1 Z)^H(P_4 Z) = 0$ for all $Z \in \mathbb{C}^{m\times n}$. Then Corollary 3.5 implies

$$\|(P_1 + P_4)Z\|_* = \|P_1 Z\|_* + \|P_4 Z\|_* \quad \forall Z \in \mathbb{C}^{m\times n}. \tag{11}$$

Also note that $\operatorname{rank}(P_k Z) \leqslant r$ for all $Z \in \mathbb{C}^{m\times n}$ and for $k = 1, 2, 3$.

Let $E = X^\star - X$ and let $P_4 E = \sum_{j\geqslant 1} \widetilde{\sigma}_j \widetilde{u}_j \widetilde{v}_j^H$ be the singular value decomposition of $P_4 E$ with singular values in decreasing order. Here $\widetilde{\sigma}_j = 0$ if $i > \operatorname{rank}(P_4 E)$. For $k \geqslant 1$, define projection operator $Q_k$ by

$$Q_k Z = \sum_{j=(k-1)r+1}^{kr} \langle Z, \widetilde{u}_j \widetilde{v}_j^H \rangle_{\mathbb{C}^{m\times n}} \widetilde{u}_j \widetilde{v}_j^H.$$

Then we have $P_4 E = \sum_{k\geqslant 1} Q_k E$ and $\operatorname{rank}(Q_k E) \leqslant r$ for all $k \geqslant 1$. Now, for all $k \geqslant 2$, we have

$$\|Q_k E\|_F \leqslant r^{1/2} \|Q_k E\|_2 \leqslant r^{-1/2} \|Q_{k-1} E\|_*$$

and therefore

$$\sum_{k\geqslant 2} \|Q_k E\|_F \leqslant r^{-1/2} \sum_{k\geqslant 1} \|Q_k E\|_* = r^{-1/2} \|P_4 E\|_*. \tag{12}$$

It follows that

$$\|P_4 E - Q_1 E\|_F = \|\sum_{k\geqslant 2} Q_k E\|_F \leqslant \sum_{k\geqslant 2} \|Q_k E\|_F \leqslant r^{-1/2} \|P_4 E\|_*. \tag{13}$$

Next, since $X^\star$ is a solution to P,

$$\begin{aligned}
\|X\|_* &\geqslant \|X^\star\|_* = \|X+E\|_* = \|(P_1+P_2+P_3+P_4)(X+E)\|_* \\
&\geqslant \|(P_1+P_4)(X+E)\|_* - \|(P_2+P_3)(X+E)\|_* \\
&= \|P_1(X+E)\|_* + \|P_4(X+E)\|_* - \|(P_2+P_3)(X+E)\|_* \\
&\geqslant \|P_1X\|_* - \|P_1E\|_* + \|P_4E\|_* - \|P_4X\|_* - \|(P_2+P_3)X\|_* - \|(P_2+P_3)E\|_* ,
\end{aligned}$$

where the equality in the third line follows from (11). Therefore

$$\|P_4E\|_* \leqslant \underbrace{\|P_1E\|_* + \|(P_2+P_3)E\|_*}_{\text{(a)}} + \underbrace{\|X\|_* - \|P_1X\|_* + \|(P_2+P_3)X\|_* + \|P_4X\|_*}_{\text{(b)}} .$$

**Lemma 3.7** *Let $\alpha > 0$ be a constant and let $x, y \in \mathbb{R}$ satisfy $x^2 + y^2 = 1$. Then $x + \alpha y \leqslant 2\alpha/\sqrt{\alpha^2+1}$.*

**Proof** Let $(x_0, y_0) = \arg\max_{(x,y)}\{x + \alpha y : \; x^2 + y^2 = 1\}$. We may assume $x_0, y_0 \geqslant 0$ and then $y_0 = \sqrt{1-x_0^2}$. Let $f(x) = x + \alpha\sqrt{1-x^2}$. Then $\frac{df(x)}{dx}\Big|_{x=x_0} = 1 + \frac{\alpha x}{\sqrt{1-x^2}} = 0$. Therefore the maximum $2\alpha/\sqrt{\alpha^2+1}$ is achieved when $x_0 = \alpha/\sqrt{\alpha^2+1}$ and $y_0 = 1/\sqrt{\alpha^2+1}$. ∎

Define a constant $\gamma \triangleq \frac{2\sqrt{2}}{\sqrt{3}}$. We further bound (a) by

$$\begin{aligned}
&\|P_1E\|_* + \|(P_2+P_3)E\|_* \\
\leqslant\; & r^{1/2}\|P_1E\|_F + (2r)^{1/2}\|(P_2+P_3)E\|_F \\
\leqslant\; & \gamma r^{1/2}\|(P_1+P_2+P_3)E\|_F ,
\end{aligned}$$

where the first inequality follows from Lemma 3.6 with $\mathrm{rank}(P_1E) \leqslant r$ and $\mathrm{rank}((P_2+P_3)E) \leqslant 2r$ and the second inequality is obtained by invoking Lemma 3.7 with $x = \|P_1E\|_F / \|(P_1+P_2+P_3)E\|_F$, $y = \|(P_2+P_3)E\|_F / \|(P_1+P_2+P_3)E\|_F$, and $\alpha = \sqrt{2}$.

Next we further bound $(b)$ by

$$
\begin{aligned}
& \|X\|_* - \|P_1 X\|_* + \|(P_2+P_3)X\|_* + \|P_4 X\|_* \\
\leqslant\ & \|P_1 X\|_* + \|P_4 X\|_* + \|(P_2+P_3)X\|_* - \|P_1 X\|_* + \|(P_2+P_3)X\|_* + \|P_4 X\|_* \\
=\ & 2\|(P_2+P_3)X\|_* + 2\|P_4 X\|_* \\
\leqslant\ & 2(2r)^{1/2}\|(P_2+P_3)X\|_F + 2r^{1/2}\|P_4 X\|_F \\
\leqslant\ & 2\gamma r^{1/2}\|(P_2+P_3+P_4)X\|_F = 2\gamma r^{1/2}\|X - X_r\|_F .
\end{aligned}
$$

Therefore

$$
\begin{aligned}
\|P_4 E\|_* &\leqslant \gamma r^{1/2}\|(P_1+P_2+P_3)E\|_F + 2\gamma r^{1/2}\|X - X_r\|_F \\
&\leqslant \gamma r^{1/2}\|(P_1+P_2+P_3)E + Q_1 E\|_F + 2\gamma r^{1/2}\|X - X_r\|_F . \qquad (14)
\end{aligned}
$$

Here we used the fact that by construction $Q_1$ is orthogonal to $P_1$, $P_2$, and $P_3$.

Combining (13) and (14), we have

$$
\|P_4 E - Q_1 E\|_F \leqslant \gamma\|(P_1+P_2+P_3)E + Q_1 E\|_F + 2\gamma\|X - X_r\|_F . \qquad (15)
$$

Next we bound $\|E - (P_4 E - Q_1 E)\|_F = \|(P_1+P_2+P_3)E + Q_1 E\|_F$. Since $\mathrm{rank}((P_1+P_2+P_3)E) \leqslant 2r$ and $\mathrm{rank}(Q_1 E) \leqslant r$, by the subadditivity of rank, $\mathrm{rank}(E - (P_4 E - Q_1 E)) \leqslant 3r$.

Since $P_4 E - Q_1 E = \sum_{k\geqslant 2} Q_k E$

$$
\begin{aligned}
\|\mathcal{A}(E - (P_4 E - Q_1 E))\|_2^2 &= \underbrace{\langle \mathcal{A}(E - (P_4 E - Q_1 E)), \mathcal{A}E\rangle_{\mathbb{C}^p}}_{\text{(c)}} \\
&- \underbrace{\langle \mathcal{A}(E - (P_4 E - Q_1 E)), \mathcal{A}\sum_{k\geqslant 2} Q_k E\rangle_{\mathbb{C}^p}}_{\text{(d)}} . \qquad (16)
\end{aligned}
$$

We bound (c) by

$$
\begin{aligned}
&\langle \mathcal{A}\left(E-(P_4E-Q_1E)\right), \mathcal{A}E\rangle_{\mathbb{C}^p} \\
\leqslant\ & \left\|\mathcal{A}\left(E-(P_4E-Q_1E)\right)\right\|_2 \left\|\mathcal{A}E\right\|_2 \\
\leqslant\ & 2\epsilon\sqrt{1+\delta_{3r}(\mathcal{A})}\left\|E-(P_4E-Q_1E)\right\|_F . \qquad (17)
\end{aligned}
$$

Here we used the rank-restricted isometry property of $\mathcal{A}$ with $\text{rank}(E-(P_4E-Q_1E))\leqslant 3r$ and the fact that

$$
\left\|\mathcal{A}E\right\|_2 = \left\|\mathcal{A}(X^\star - X)\right\|_2 \leqslant \left\|b-\mathcal{A}X^\star\right\|_2 + \left\|b-\mathcal{A}X\right\|_2 \leqslant 2\epsilon,
$$

because $X^\star$ is a solution to P. Next we bound (d). For each $k\geqslant 2$,

$$
\begin{aligned}
&\left|\langle \mathcal{A}\left(E-(P_4E-Q_1E)\right), \mathcal{A}Q_kE\rangle_{\mathbb{C}^p}\right| \\
=\ & \left|\langle \mathcal{A}\left((P_1+P_2+P_3)E+Q_1E\right), \mathcal{A}Q_kE\rangle_{\mathbb{C}^p}\right| \\
\leqslant\ & \left|\langle \mathcal{A}(P_1E+Q_1E), \mathcal{A}Q_kE\rangle_{\mathbb{C}^p}\right| + \left|\langle \mathcal{A}(P_2+P_3)E, \mathcal{A}Q_kE\rangle_{\mathbb{C}^p}\right| \\
\leqslant\ & \delta_{3r}(\mathcal{A})\left\|P_1E+Q_1E\right\|_F\left\|Q_kE\right\|_F + \delta_{3r}(\mathcal{A})\left\|(P_2+P_3)E\right\|_F\left\|Q_kE\right\|_F \\
\leqslant\ & \sqrt{2}\delta_{3r}(\mathcal{A})\left\|(P_1+P_2+P_3)E+Q_1E\right\|_F\left\|Q_kE\right\|_F \\
=\ & \sqrt{2}\delta_{3r}(\mathcal{A})\left\|E-(P_4E-Q_1E)\right\|_F\left\|Q_kE\right\|_F , \qquad (18)
\end{aligned}
$$

where the second inequality follows from Proposition 3.3 because $Q_kP_j = 0$, $Q_kQ_1 = 0$ for $j = 1,2,3$ and $k\geqslant 2$ and these projections are defined by pairwise orthogonal rank-one matrices.

Applying (17) and (18) to (16), we have

$$\begin{aligned}
& \left\|\mathcal{A}\left(E-(P_4E-Q_1E)\right)\right\|_2^2 \\
\leqslant\ & \left\|E-(P_4E-Q_1E)\right\|_F\left(2\epsilon\sqrt{1+\delta_{3r}(\mathcal{A})}+\sqrt{2}\delta_{3r}(\mathcal{A})\sum_{k\geqslant 2}\left\|Q_kE\right\|_F\right) \\
\leqslant\ & \left\|E-(P_4E-Q_1E)\right\|_F\left(2\epsilon\sqrt{1+\delta_{3r}(\mathcal{A})}+\sqrt{2}\delta_{3r}(\mathcal{A})r^{-1/2}\left\|P_4E\right\|_*\right), \quad (19)
\end{aligned}$$

where the second inequality follows from (12).

From the rank-restricted isometry property of $\mathcal{A}$,

$$\left\|\mathcal{A}\left(E-(P_4E-Q_1E)\right)\right\|_2^2 \geqslant (1-\delta_{3r}(\mathcal{A}))\left\|E-(P_4E-Q_1E)\right\|_F^2. \quad (20)$$

Combining (19) and (20), we obtain

$$\left\|E-(P_4E-Q_1E)\right\|_F \leqslant \alpha\epsilon+\rho r^{-1/2}\left\|P_4E\right\|_*,$$

where

$$\begin{aligned}
\alpha &= \frac{2\sqrt{1+\delta_{3r}(\mathcal{A})}}{1-\delta_{3r}(\mathcal{A})} \\
\rho &= \frac{\sqrt{2}\delta_{3r}(\mathcal{A})}{1-\delta_{3r}(\mathcal{A})}.
\end{aligned}$$

Using (14) and the fact that $Q_1$ is orthogonal to $P_1$, $P_2$, and $P_3$,

$$\begin{aligned}
\left\|E-(P_4E-Q_1E)\right\|_F &\leqslant \alpha\epsilon+\gamma\rho\left\|(P_1+P_2+P_3)E\right\|_F+2\gamma\rho\left\|X-X_r\right\|_F \\
&\leqslant \alpha\epsilon+\gamma\rho\left\|(P_1+P_2+P_3)E+Q_1E\right\|_F+2\gamma\rho\left\|X-X_r\right\|_F \\
&= \alpha\epsilon+\gamma\rho\left\|E-(P_4E-Q_1E)\right\|_F+2\gamma\rho\left\|X-X_r\right\|_F.
\end{aligned}$$

To proceed we use the assumption $\delta_{3r}(\mathcal{A})<\frac{1}{1+4/\sqrt{3}}$, which implies $1-\gamma\rho >$

0, hence

$$\|E - (P_4E - Q_1E)\|_F \leqslant (1 - \gamma\rho)^{-1} (\alpha\epsilon + 2\gamma\rho \|X - X_r\|_F) .$$

Finally,

$$\begin{aligned} \|E\|_F &\leqslant \|E - (P_4E - Q_1E)\|_F + \|P_4E - Q_1E\|_F \\ &\leqslant (1 + \gamma) \|E - (P_4E - Q_1E)\|_F + 2\gamma \|X - X_r\|_F \\ &\leqslant (1 - \gamma\rho)^{-1} [(1 + \gamma)\alpha\epsilon + 2\gamma(1 + \rho) \|X - X_r\|_F] , \end{aligned}$$

where the second inequality follows from (15). ∎

# 4 Conclusion

In this paper, we derived an extended performance guarantee of nuclear norm minimization with an ellipsoidal constraint. Unlike existing performance guarantee, this constraint accommodates problem formulation in which the matrix is only approximately low rank, or in which there is noise in the measurements. The condition for the performance guarantee is given in terms of the rank-restricted isometry property of the linear operator in the constraint. The new performance guarantee in this paper ensures the quality of a low-rank approximate solution obtained by nuclear norm minimization with an ellipsoidal constraint. Such an approximate solution can be found by using existing polynomial-time algorithms.

# References

[Can08] E.J. Candes. The restricted isometry property and its implications for compressed sensing. *Comptes rendus-Mathématique*, 346(9-10):589–592, 2008.

[CCS08] J.F. Cai, E.J. Candes, and Z. Shen. A singular value thresholding algorithm for matrix completion. *Arxiv preprint arXiv:0810.3286*, 2008.

[CT05] E.J. Candes and T. Tao. Decoding by linear programming. *IEEE Transactions on Information Theory*, 51(12):4203–4215, 2005.

[Faz02] M. Fazel. *Matrix rank minimization with applications.* PhD thesis, Stanford University, 2002.

[FHB01] M. Fazel, H. Hindi, and S.P. Boyd. A rank minimization heuristic with application to minimum order system approximation. In *Proceedings American Control Conference*, volume 6, pages 4734–4739, 2001.

[LLR95] N. Linial, E. London, and Y. Rabinovich. The geometry of graphs and some of its algorithmic applications. *Combinatorica*, 15(2):215–245, 1995.

[RFP07] B. Recht, M. Fazel, and P.A. Parrilo. Guaranteed minimum-rank solutions of linear matrix equations via nuclear norm minimization. *Arxiv preprint arXiv:0706.4138*, 2007.